Growth, electrical, structural, and magnetic properties of half-Heusler CoTi$_{1-x}$Fe$_x$Sb

S. D. Harrington[1], A. D. Rice[1], T. Brown-Heft[1], B. Bonef[1], A. Sharan[6], A.P. McFadden[2], J. A. Logan[1], M. Pendharkar[2], M.M. Feldman[3] O. Mercan[4], A. G. Petukhov[5], A. Janotti[6,7] L. Çolakerol Arslan[4], and C. J. Palmstrøm[1,2]*

[1]Materials Department, University of California, Santa Barbara, CA 93106, USA
[2]Department of Electrical & Computer Engineering, University of California, Santa Barbara, CA 93106, USA
[3] Physics Department, University of California, Santa Barbara, CA 93106, USA
[4] Physics Department, Gebze Technical University, Kocaeli, 41400 Turkey
[5] Physics Department, South Dakota School of Mines and Technology, Rapid City, SD 57701, USA
[6]Department of Physics and Astronomy, University of Delaware, Newark, DE 19716, USA
[7]Department of Materials Science & Engineering, University of Delaware, Newark, DE 19716, USA

*Correspondence to: cpalmstrom@ece.ucsb.edu

**Abstract**

Epitaxial thin films of the substitutionally alloyed half-Heusler series CoTi$_{1-x}$Fe$_x$Sb were grown by molecular beam epitaxy on InAlAs/InP(001) substrates for concentrations $0.0 \leq x \leq 1.0$. The influence of Fe on the structural, electronic, and magnetic properties was studied and compared to that expected from density functional theory. The films are epitaxial and single crystalline, as measured by reflection high-energy electron diffraction and X-ray diffraction. Using *in-situ* X-ray photoelectron spectroscopy, only small changes in the valence band are detected for $x \leq 0.5$. For films with $x \geq 0.05$, ferromagnetism is observed in SQUID magnetometry with a saturation magnetization that scales linearly with Fe content. A dramatic decrease in the magnetic moment per formula unit occurs when the Fe is substitutionally alloyed on the Co site indicating a strong dependence on the magnetic moment with site occupancy. A cross over from both in-plane and out-of-plane magnetic moments to only in-plane moment occurs for higher concentrations of Fe. Ferromagnetic resonance indicates a transition from weak to strong interaction with a reduction in inhomogeneous broadening as Fe content is increased. Temperature dependent transport reveals a semiconductor to metal transition with thermally activated behavior for $x \leq 0.5$. Anomalous Hall effect and large negative magnetoresistance (up to -18.5% at 100 kOe for x=0.3) are observed for higher Fe content films. Evidence of superparamagnetism for x=0.3 and x=0.2 suggests for moderate levels of Fe, demixing of the CoTi$_{1-x}$Fe$_x$Sb films into Fe rich and Fe deficient regions may be present. Atom probe tomography is used to examine the Fe distribution in a x=0.3 film. Statistical analysis reveals a nonhomogeneous distribution of Fe atoms throughout the film, which is used to explain the observed magnetic and electrical behavior.

I-    Introduction

Half-Heusler (h-H) and full-Heusler (f-H) compounds are an exciting class of ternary intermetallics due to their diverse electrical and magnetic properties, including semiconducting[1], half-metallic[2], and thermoelectric[3]. Additionally, topologically non-trivial behavior has been predicted[4] and recently observed[5] within these compounds. The possibility to generate highly spin-polarized currents from either the half-metallic or topologically non-trivial variants have made Heusler compounds attractive for many spintronic applications. The electronic and magnetic properties have a strong dependence on the number of valence electrons per formula unit (f.u.) and chemical composition[6]. They can be alloyed in a similar manner to compound semiconductors leading to controllable tuning of these electronic and magnetic properties[7–10]. Using this technique, promising thermoelectric performance has already been shown for h-H based thermoelectrics[7] as well as Fermi level tuning in f-H based magnetic tunnel junctions (MTJs) where room temperature tunneling magnetoresistance of greater than 100% has been reported[8–10]. Additionally, the h-H crystal structure ($C1_b$) and their lattice parameters are closely related to III-V compound semiconductors, suggesting the possibility of h-H/III-V heterostructures with unique properties.

A number of h-H compounds with 18 valence electrons per f.u. have been calculated to exhibit semiconducting properties[11]. In particular, CoTiSb has previously been studied for its thermoelectric properties[12] and is predicted to be a semiconductor with an indirect bandgap ($\Gamma$-X) of ~1 eV[13,14]. Prior work on the MBE growth of CoTiSb films on $In_{0.52}Al_{0.48}As$ buffer layers obtained relatively low carrier concentrations (~$5 \times 10^{17}$ $cm^{-3}$) and mobilities (~500 $cm^2$/V-s) at room temperature comparable to Si with the same carrier concentrations[15]. Additionally, CoTiSb has a lattice parameter of 5.88 Å, which is nearly lattice matched to InP. By substitutionally alloying Fe into the Ti site, a net gain of four valence electrons per f.u. is obtained, and CoTiSb undergoes a transition from a non-magnetic semiconductor to a ferromagnet[16]. A dilute magnetic semiconductor with a high Curie temperature was reported for low levels of Fe alloying in bulk single crystals[17], which could be useful for spintronic applications such as MTJs or spin injection into nonmagnetic materials. However, for many device applications thin films are a necessary requirement. Previous work by Sun *et al.*[18] reported the growth and properties of Fe alloyed CoTiSb epitaxial thin films prepared by the magnetron sputtering method. For the intrinsic CoTiSb thin films, carrier concentrations of ~$10^{21}$ $cm^{-3}$ and mobilities of ~2 $cm^2$/V-s at room temperature were reported, which was attributed to Ti-deficiency. In addition, because the Fe was introduced by adding Fe flakes to the CoTiSb target, the composition and stoichiometry of the film could not be closely controlled. In the present study, epitaxial thin films of Fe alloyed CoTiSb are grown by molecular beam epitaxy (MBE), which allows for the precise control of stoichiometry. The growth, structural, magnetic, and transport properties of the resulting films and their dependence on Fe content are discussed and compared to our own calculations based on density functional theory (DFT).

**II-    Experimental Details**

$CoTi_{1-x}Fe_xSb$ samples were grown in a VG V80 MBE system on nearly lattice-matched unintentionally doped $In_{0.52}Al_{0.48}As$ (referred to as InAlAs) buffer layers epitaxial grown on semi-insulating InP:Fe (001) substrates. The InAlAs layers were 400 nm thick, grown in a separate conventional III-V MBE system and then arsenic capped and transferred through air into a dedicated metals MBE system for growth of the $CoTi_{1-x}Fe_xSb$ layers. After the samples were reintroduced to ultra-high vacuum (UHV), the arsenic cap was desorbed to reveal the As-terminated (2×4)/c(2×8) InAlAs surface. CoTi$_{1-x}$Fe$_x$Sb thin films were grown by simultaneous evaporation of Co, Ti, Fe, and Sb using stoichiometric fluxes with a total flux of 9x10$^{16}$ atoms/cm$^2$·hr, giving an approximate growth rate of 2.5 Å/min. 9x10$^{16}$ atoms/cm$^2$ were deposited for each film which corresponds to approximately 15 nm for pure CoTiSb. All fluxes were calibrated *ex situ* by measuring the elemental atomic areal density of calibration sample layers grown on Si substrates using Rutherford backscattering spectrometry (RBS). Samples were grown at temperatures in the range 200-380°C as measured by a thermocouple that is calibrated to the arsenic desorption temperature of arsenic capped GaAs[19]. The surface was monitored *in situ* during growth using reflection high-energy electron diffraction (RHEED). Following growth, samples were cooled down before UHV transfer to an e-beam evaporator for room temperature deposition of a ~10 nm amorphous AlO$_x$ protective capping layer.

The crystal structure and magnetic properties were analyzed *ex situ* using X-ray diffraction (XRD), superconducting quantum interference device (SQUID) magnetometry, and ferromagnetic resonance (FMR). SQUID magnetometry field sweeps were conducted at 5K following a 5000 Oe-field cooldown with separate sweeps with the applied field along the [110], [100], and [001] (out-of-plane) crystallographic directions. Magnetic moment vs temperature data was collected with a 100 Oe field applied from 5 to 400 K. The FMR spectra were collected on a conventional x-band (ν=9.8GHz) Joel-FA300 spectrometer. The measurements were performed in 110–350 K temperature range using a Joel LN$_2$ flow cryostat. Electrical characterization was performed between 2 and 300 K in a He-4 cryostat using a standard dc technique in a L-shaped Hall bar geometry of length 1mm (longitudinal voltage leads were spaced 200 μm apart and 150 μm wide) aligned along the [110] and [$\bar{1}$10] directions. Hall bars were fabricated using contact lithography and Ar ion milling with e-beam deposited Ti/Au contacts. R vs H measurements were performed with a constant applied current while measuring V$_{xx}$ and V$_{xy}$ so as to determine R$_{xx}$ and R$_{xy}$ components simultaneously during the out-of-plane magnetic field sweep from 100 kOe to -100 kOe and back to 100 kOe.

The magnetization and density of states were calculated using DFT[20,21] with the revised Perdew-Burke-Ernzerhof functional for solids for exchange and correlation (PBEsol)[22] as implemented in VASP code[23,24]. The interactions between the valence electrons and the ionic cores are treated using projector-augmented wave potentials[25,26]. The random alloy structures, for varying Fe concentration, were generated using special quasi-random structures (SQS)[27] based on a 90-atom supercell that were determined with the Alloy Theoretic Automated Toolkit (ATAT) code[28]. SQS has been successful in describing the electronic and thermodynamic properties of various disordered systems[29–31]. The calculations are performed using an energy cutoff of 350 eV for plane-wave basis set expansion and a grid of Γ-centered 6x6x6 *k*-points in reciprocal space for integrations over the Brillouin zone.

## III- Results and Discussion
### A) Surface, structural, and electronic characterization

During growth, a (2x1) surface reconstruction was inferred from bright, streaky RHEED patterns for x≤0.5 (Fig 1), similar to that observed in intrinsic CoTiSb[15,32]. For the pure CoFeSb film, the RHEED pattern consisted of faint streaks as well as bulk diffraction spots indicating roughening of the surface and lower film quality. To minimize interfacial reactions and phase segregation, a lower growth temperature was necessary as the Fe content was increased. This is similar to low-temperature MBE required to achieve (Ga,Mn)As thin films[33]. However, polycrystalline rings were observed in RHEED below 200°C growth temperature. Therefore, for the highest Fe content film, an optimal growth temperature of 200°C was used to maintain single crystal growth of CoFeSb.

Figure 2 shows XRD 2θ-ω scans for $CoTi_{1-x}Fe_xSb$ films for x=0.0, 0.2, 0.3, 0.5, and 1.0 grown on InAlAs/InP (001). The sharp peaks at 2θ=30.44° and 63.34° correspond to the InP (002) and (004) substrate reflections, respectively. The $CoTi_{1-x}Fe_xSb$ and InAlAs (002) and (004) peaks are nearly overlaid on the InP peaks indicating the close lattice match. Other than the (00$l$) peaks and thickness fringes, no additional peaks in the XRD scans are observed. Figure 2(b) shows a scan centered around the (004) reflection. Here, finite thickness fringes can be clearly resolved for x≤0.5 corresponding to a thickness of 15.6, 14.4, 14.3, and 13.3 nm for x=0.0, 0.2, 0.3, and 0.5 respectively, in good agreement with the film thickness expected from the RBS calibrations. These fringes indicate a smooth, abrupt interface between the $CoTi_{1-x}Fe_xSb$ and InAlAs for up to x=0.5. The additional large peak observed in each scan corresponds to the InAlAs buffer layer. Small deviations from the intended composition of $In_{0.52}Al_{0.48}As$ led to variations of the lattice parameter. The broad peak to the left for x≤0.5 and to the right for x=1.0 of InP are from the $CoTi_{1-x}Fe_xSb$ films. A small increase in the out-of-plane lattice parameter from 5.88 Å for pure CoTiSb is observed as Fe content is first increased with a dramatic decrease for the pure CoFeSb (5.81 Å). The lattice parameter of CoFeSb agrees with the theoretically predicted value of 5.81 Å[34]. The observed lattice parameter bowing can partially be attributed to slightly different strain conditions due to variation in the buffer lattice parameter. These XRD patterns combined with the RHEED images, indicate an epitaxial cube-on-cube growth with no detectable secondary phases or orientations and are suggestive of abrupt interfaces and high crystalline quality for the films with lower iron content.

The effect of Fe on the CoTiSb electronic structure was probed by performing *in-situ* XPS on intrinsic CoTiSb, $CoTi_{0.8}Fe_{0.2}Sb$, and $CoTi_{0.5}Fe_{0.5}Sb$ and compared to DFT calculated density of states (DOS). The calculated DOS for $CoTi_{1-x}Fe_xSb$ for $x$ = 0.0, 0.2 and 0.5 is shown in Fig. 3(a). As expected, we find pure CoTiSb to be a semiconductor, and $CoTi_{1-x}Fe_xSb$ with $x$=0.2 and 0.5 to be metallic. The shape of the DOS in the valence band for $x$ = 0.2 and 0.5 follows that of the pure CoTiSb, since these are composed mostly of Co and Sb orbitals. The bands above the Fermi level changes for $x$=0.5 compared to pure CoTiSb, which is expected since the conduction band (up to ~3 eV) has major contributions from the Fe and/or Ti atoms. The normalized valence band spectra excited by Al $K_{α1}$ radiation are shown in Fig. 3(b). For the CoTiSb spectrum, good agreement is observed with the DOS and resembles previously reported spectra at similar excitation energies[16,18,32,35]. For the x=0.2 and x=0.5 films, the spectra show similar structure to that of CoTiSb with only a few small binding energy shifts, consistent with our calculated DOS. Here good agreement between the calculated DOS and the measured valence band spectrum is observed, suggesting no additional, non-h-H phase is present.

**B) Magnetic properties**

The magnetic properties of the alloy series were studied using *ex-situ* SQUID magnetometry and FMR. From SQUID magnetometry, ferromagnetic order was observed for CoTi$_{1-x}$Fe$_x$Sb films with x≥0.05 within the detection limit of the measurement. In-plane hysteresis loops with the applied field oriented along the [110] direction taken at 5K are shown in Fig. 4(a). The saturation magnetization of the films increases with increasing Fe content, which was used to calculate the net contribution of an iron atom to the total magnetic moment per f.u. in units of Bohr magneton ($\mu_B$). Additionally, a decrease in the coercive field from 500 to 10 Oe for compositions from x=0.1 to x=1.0 can be observed. The magnetic moment per f.u. is plotted in Fig. 4(b). A linear dependence of 3.9 $\mu_B$/Fe atom is observed up to x=0.5, close to the Slater-Pauling expected 4 $\mu_B$/Fe atom-f.u[17]. However, including pure CoFeSb, which displays 3.2 $\mu_B$/f.u., gives an overall linear fit of 3.3 $\mu_B$/Fe atom-f.u. This deviation from the well-behaved linear fit observed at low iron concentration may be caused by poor crystal quality for the CoFeSb sample. The error bars are from differences in saturation magnetization measured for the distinct crystallographic directions. These differences arise due to the finite sample size effects on the SQUID pickup coils[36].

Previously, Kroth *et al.* studied bulk CoTi$_{1-x}$Fe$_x$Sb crystals with x=0.05 and 0.1 and obtained m=3.5$\mu_B$/Fe atom-f.u and m=3.7 $\mu_B$/Fe atom, respectively[17] consistent with the results reported here. Sun *et. al.* investigated Fe doped CoTiSb films on MgO (001) up to 37% doping ratio and obtained m=3 $\mu_B$/f.u for 20 nm thick films. The deviations from the expected value of 4 $\mu_B$/Fe atom have been attributed to disorder on the site occupancy of the Fe atom. It was suggested that Fe atoms occupying alternative Wyckoff positions are expected to contribute significantly less magnetic moment[17]. Thus the discrepancies on the measured bulk spin moment per Fe atom may be correlated with the difficulty in preparing well-ordered and stoichiometric films. For example, FeTiSb would be expected to have -1 $\mu_B$/f.u. from the simplified m = N$_V$-18 Slater-Pauling curve. Thus Fe occupying other sites would be expected to contribute less magnetic moment. To verify this, the substitutional series Co$_{1-y}$Fe$_y$TiSb was grown with the intent of replacing Co with Fe, and the magnetic properties measured. The saturation magnetic moments for films with y=0.1, y=0.2, and y=0.5 are plotted in Fig. 4(b). Again, a linear dependence with Fe content is found; however, a drastically reduced net magnetic moment of 0.42 $\mu_B$/Fe atom is observed. This discrepancy is likely due to disorder within the films.

To understand the effects of Fe alloying and disorder, the total magnetization of CoTi$_{1-x}$Fe$_x$Sb and Co$_{1-y}$Fe$_y$TiSb atomically resolved moments were calculated using DFT for CoTi$_{1-x}$Fe$_x$Sb and Co$_{1-y}$Fe$_y$TiSb, as well as several structures with mixed alloying or antisite swap defects. When Fe atoms substitute on the Ti sites in CoTiSb, the total magnetization of the alloy increases by 3.92 $\mu_B$ per Fe atom, as shown in Fig 5, though the magnetization in pure CoFeSb drops to 3.84 $\mu_B$. The observed magnetic moment of about 4 $\mu_B$ per Fe atom in the CoTi$_{1-x}$Fe$_x$Sb alloys arises from strong *d-d* coupling between Co and Fe *d* orbitals and spin splitting, with ~3 $\mu_B$ centered on Fe atoms and ~1 $\mu_B$ centered on Co atoms, for all Fe concentrations.

Alternatively, when Fe is substituted on the Co sites in Co$_{1-y}$Fe$_y$TiSb, there is an expected reduction in the overall net magnetization due to the Fe atom and it has the opposite sign to that for Fe on Ti sites, CoTi$_{1-x}$Fe$_x$Sb. From DFT calculations summarized in Fig 5, the total magnetization per Fe atoms jumps to -0.84 $\mu_B$ for x=0.1 and then saturates to -0.90 $\mu_B$ for higher Fe concentrations. The projected density of states on Co, Fe and Ti *d* orbitals are shown in Fig 6(b) for the Co$_{0.5}$Fe$_{0.5}$TiSb alloy. In this case, as well as lower Fe compositions, most of the magnetization is concentrated on the Fe atoms with negligible contributions from the neighboring Co or Ti atoms. This is consistent with the previous results[34]. Note that the magnetic moment observed in this case has the opposite sign of that calculated in the case of CoTi$_{1-x}$Fe$_x$Sb. To check if this sign reversal is preserved in the case of mixed site alloying we calculate magnetization in Co$_{1-y}$Ti$_{1-x}$Fe$_{x+y}$Sb (x=y) for different Fe concentrations, in which half of the Fe atoms were substituted on the Co site and another half on Ti site. In this case, we observe net magnetic moment of 1.5 $\mu_B$/Fe atom, consistent with the Slater-Pauling rule. The magnetic moments from Fe on Co (Fe$_{Co}$) and Fe on Ti (Fe$_{Ti}$) sites partially compensate each other, in agreement with the sign reversal observed for the two ordered alloy systems. Finally, we also calculated the moment for the alloy series with complete Co$_{Ti}$ disorder(Co$_{1-y}$Fe$_y$)(Ti$_{1-x}$Co$_x$)Sb (x=y), where Fe atoms substitute on the Co site and the displaced Co occupies the induced vacancies in the Ti site. Interestingly, the calculated magnetic moment in this case is ~2.38 $\mu_B$/Fe atom, which deviates significantly from the magnetic moment expected from Slater Pauling rule of 4.0 $\mu_B$/Fe atom. This disorder induced deviation from Slater-Pauling is consistent with that predicted in NiMnSb, where even a few percent disorder reduced the net magnetization[37]. The magnetic moment calculated for different stoichiometry/disorders arising from Fe atoms occupying different sites are summarized in Table I.

Therefore, the observed magnetic moment of 0.42 $\mu_B$/Fe atom for the substitutional series Co$_{1-y}$Fe$_y$TiSb can be explained as arising from disorder in occupation of Fe atoms, i.e., having a mix of Fe$_{Co}$ and Fe$_{Ti}$ sites. Due to the sign ambiguity in experimentally measured magnetic moments, two values are possible the site ordering ratio in this system. For a positive magnetic moment of 0.42 $\mu_B$/Fe, we predict a ratio of Fe$_{Co}$/Fe$_{Ti}$ ≈ 2.6, while a negative magnetic moment would give a ratio of Fe$_{Co}$/Fe$_{Ti}$ ≈ 9. For the CoTi$_{1-x}$Fe$_x$Sb films, a higher magnetic moment is indicative of better ordering within the film of Fe occupying the Ti site. Thus the observed trend of 3.9 $\mu_B$/Fe atom for lower Fe content films suggests that the MBE prepared thin films are well ordered with the majority of Fe atoms occupying the Ti site.

In addition to site disorder calculations, spin resolved density of states was calculated for select concentrations of the ordered alloys. The projected density of states on Co, Ti and Fe *d* orbitals are shown in Fig 6(a) for the CoTi$_{0.5}$Fe$_{0.5}$Sb alloy. The spin down density of states is zero with non-zero spin up density of states at Fermi level, suggesting the alloy to be half-metallic. In fact, we found that alloys with up to x=0.5 are half-metallic.

To better understand the magnetic switching behavior and anisotropy in the $CoTi_{1-x}Fe_xSb$ system, a detailed SQUID analysis was performed. Figure 7 shows hysteresis loops for the x=0.3 and x=0.5 samples with the applied field along the [110], [100], and [001] crystallographic directions. The [110] and [100] correspond to the field in the plane of the film, while the [001] is out of plane. For the x=0.3 film (Fig. 7a), remanence can be observed in all three directions with the largest coercive field ($H_c$=600 Oe) observed in the [001] direction. Additionally, there is no clear easy axis. This suggests that the magnetic moments on the Fe atoms are only weakly coupled. In contrast, for the x=0.5 film (Fig. 7b) only the in plane ([110] and [100]) directions show clear remanence with an easy axis along the [110] direction. The small remanence observed in the [001] direction can be attributed to small misalignment of the sample in the SQUID sample tube and reflects a small in-plane component contributing to the signal. The difference in crystallographic dependence between the two samples indicates a competition between the magnetic anisotropy terms for the different composition films. XRD reciprocal space maps reveal the films do not possess a significant tetragonal distortion, and magnetocrystalline anisotropy would not be expected to contribute to perpendicular anisotropy in cubic Heusler crystals. Surface or interface anisotropy typically depends on the interface chemical bonding (e.g. CoFeB/MgO)[38] but may also depend on interfacial strain. Changing lattice constants as a function of Fe content, with resulting changes in strain condition, could explain the perpendicular spin reorientation transition, however such changes in strain condition were small. Films with high magnetic moment per volume have a correspondingly large shape anisotropy, which tends to confine magnetization into the plane of the film. As the moment per volume of a thin film decreases, shape anisotropy decreases, allowing any surface anisotropy present to dominate. We speculate that for x=0.3, shape anisotropy is sufficiently low that interface anisotropy dominates, but is large enough for x=0.5 to confine the magnetization in the film plane. For lower Fe content films, the strength of magnetic interactions are too small to produce strong magnetic ordering and anisotropy.

The temperature dependence of the magnetic moment can be seen in the insets of Fig. 7. For the film with x=0.3, a sharp decrease in the magnetic moment is observed upon warming, with a dramatic change in slope around 50 K, whereupon a more gradual decrease is observed. The large change in slope is suggestive of a magnetic phase transition. The nature of this possible phase transition will be discussed in section D. In contrast, the x=0.5 sample shows a Curie-Weiss like temperature dependence, with a gradual decrease in the magnetic moment for increasing temperature and a Curie temperature beyond 400 K.

The magnetic anisotropy and microscopic nature of the films were investigated further by means of FMR. FMR spectra were measured in two different rotation planes. The in-plane FMR spectra for the $CoTi_{1-x}Fe_xSb$ alloy series is shown in Fig. 8(a), for which the resonance field can be determined as the field where dP/dH=0 line cuts the dP/dH vs H curve. Consistent with the magnetization measurements, we observe a strong enhancement in the intensity of the FMR spectra as the Fe content increases. In addition, a shift of the resonance positions to lower fields, and decrease of the peak-to-peak line width is observed with increasing Fe concentrations. The decrease of average resonance field with increasing Fe concentration indicates the enhancement of internal field and inter-particle (exchange) interactions due to closer possible distances between magnetic ions. The broadening of the linewidth with decreasing Fe content is due to the presence of the non-homogeneous local magnetic field, which modifies the resonance field as well as the line shape of the signal. The non-homogeneous local magnetic moment can be understood to arise from non-uniform distribution of Fe atoms within the films and will be discussed in section E for a x=0.3 film.

In Fig. 8(b), the angular dependence of the in-plane resonance fields at 110 K for $CoTi_{1-x}Fe_xSb$ films is presented. The value of resonance field oscillates as a function of the angle of the applied magnetic field. Since the resonance field is proportional to the effective magnetic field, the maximum (minimum) value is found when the field is parallel to the hard (easy) axis. The angular dependence of the resonant frequency immediately shows that the easy axis is along the [110] direction, consistent with SQUID hysteresis loops. For the samples with x<0.15 (not shown), the FMR signal is very weak and there is almost no shift in resonance field with angle, confirming the absence of in-plane anisotropy as measured in SQUID. $CoTi_{1-x}Fe_xSb$ film with x=0.2 is dominated by a four-fold symmetry, arising from the cubic anisotropy contribution from the cubic bulk $C1_b$ structure of CoTiSb film. The increase in Fe concentration to x=0.3 induces a strong planar two-fold uniaxial anisotropy field and again a very small fourfold in-plane anisotropy field. The origin of this behavior is consistent with two different magnetic environments for Fe. While the cubic anisotropy is associated with the ferromagnetic interactions of Fe atoms in Fe-rich regions, strong uniaxial anisotropy is due to the symmetry of the substrate. This is similar to Fe grown on GaAs where uniaxial anisotropy was observed in thin Fe films[39]. The easy axis aligning along the [110] direction, has been suggested to originate from the arsenic-bond direction on the GaAs(001) surface[40]. For the $CoTi_{0.5}Fe_{0.5}Sb$ film, the FMR signal is pronounced and an in-plane uniaxial magnetic anisotropy is observed, associated with the dipolar interactions of Fe atoms. For pure CoFeSb, cubic anisotropy is observed with the in-plane easy axes parallel to the [110] and [$\bar{1}$10] directions. The four-fold symmetry contribution can be attributed to the bulk cubic symmetry associated with the h-H crystal structure, which for the highest magnetic moment film appears to dominate.

Figure 8(c) depicts the dependence of resonance field on the polar angle with the out-of-plane configuration at 110 K. For all Fe content, the value of resonance field is minimum when the applied field is along the film plane and reaches maximum when along the film normal. As the Fe concentration increases, the resonance field is increased slightly at angles around film plane, while reduced significantly at low angle near the film normal. The perpendicular magnetic anisotropy can be associated with the epitaxial relationship between the film and the substrate in addition to the shape anisotropy.

## C) Magnetotransport

The addition of Fe is expected to have a strong effect on the electronic properties. Hence, transport measurements were performed to determine the electrical properties of the two sample series. Figure 9 shows the longitudinal sheet resistance for the $CoTi_{1-x}Fe_xSb$ thin films dependence on Fe concentration and temperature. Figure 9(a) highlights that both the room temperature and low temperature sheet resistance decrease with increasing Fe content, consistent with the increased electron concentration expected. From Fig 9(b), it can be seen that for the lower Fe alloying, the film exhibits semiconducting-like transport and thermally activated behavior. From the 1/T dependence near room temperature, activation energies of 9 meV, 7 meV, 4 meV, and 3 meV are extracted for x=0.0, 0.2, 0.3, and 0.5, respectively. As the Fe concentration increases, the temperature dependence becomes weaker until the Fe composition is greater than x=0.5, where metallic transport was observed. A high residual resistivity and thermally activated transport has been observed within other predicted half-metallic f-H compounds[41,42] and may originate from disorder within a half-metallic system. The observed temperature dependence is consistent with previous reports of sputtered films, where thermally activated behavior was observed for films up to 37 atomic % of Fe[18].

Longitudinal magnetoresistance (MR)-*H* curves for the x=0.3 and x=0.5 films are shown in Fig. 10 for temperatures between 5 K and 300 K, where the magnetic field is perpendicular to the film plane. MR is defined as

$$MR = (R_{XXH} - R_{XX0})/R_{XX0}, \qquad (1).$$

where $R_{XXH}$ and $R_{XX0}$ refer to the resistance measured with and without an applied magnetic field, respectively. For the x=0.3 film (Fig. 10a), a negative MR, with a magnitude which monotonically increases with decreasing temperature for all field strengths, is observed. The temperature dependence of the MR at 100 kOe is displayed in the inset. Here it can be seen that the rate of increase in the MR magnitude drastically changes around 40 K. This is consistent with the temperature dependence of the magnetization shown in Fig. 7(a). In addition, a relatively large MR of 18.5% is observed at 100 kOe at 5 K which does not saturate within 140 kOe. This field strength is much beyond the saturation field seen in SQUID, indicating the large MR cannot be ascribed to only the magnetization of the film. This large MR can be attributed to the suppression of spin disorder within the system, which is consistent with the enhancement of spin-dependent scattering at lower temperatures. In contrast, the MR of the x=0.5 film displays a much weaker temperature dependence. Although a negative MR is observed at higher fields, a positive MR can be observed at low field. The peak in the MR corresponds to the saturation magnetization observed in SQUID and can be attributed to anisotropic MR. The magnitude and field position of the peak decrease with increasing temperature until it disappears by 300 K.

A large anomalous Hall effect (AHE), shown in Fig 11(a-b), was observed in both the x=0.3 and x=0.5 films. The Hall resistance $R_{Hall}$ in magnetic materials can be expressed as,

$$R_H = (R_0/d)B + (R_S/d)M, \qquad (2).$$

where $R_0$ is the ordinary Hall coefficient, $d$ is the sample thickness, $R_S$ the anomalous Hall coefficient, and $M$ the magnetization of the samples. Although the contribution of the ordinary Hall effect was small in comparison to the AHE, electron dominated transport could be observed in both samples consistent with Fe being an electron donor when occupying the Ti site. Magnetic remanence was observed at low temperature for the x=0.3 film in both the MR and AHE and is highlighted in the lower inset of Fig. 11(a). The magnitude of the remnant field decreases with increasing temperature and disappears by 80 K as shown in the upper inset of Fig. 10(a). This is consistent with the out-of-plane magnetic moment observed in SQUID magnetometry for the x=0.3 film. No remanence was observed for the x=0.5 film, confirming that the magnetic moment has no out-of-plane component.

The anomalous Hall conductivity, which is given by

$$\sigma_{AH} = \frac{\rho_{AH}}{\rho_{XX}^2 + \rho_{AH}^2} \cong \frac{\rho_{AH}}{\rho_{XX}^2} \qquad (3)$$

where $\rho_{AH}$ and $\rho_{XX}$ the anomalous Hall and longitudinal resistivity respectively, exhibits a non-monotonic dependence on temperature shown in Fig.11(c,d). A minimum around 100 K is observed for both samples indicating a competition of skew scattering, side jump, and intrinsic mechanism (Berry curvatures) at these lower temperatures[43]. This is similar to that observed in magnetron sputtered 21 atomic % Fe alloyed CoTiSb thin films, which required the expanded scaling first introduced by Tian et al[44] to describe the dependence of anomalous Hall resistivity on longitudinal resistivity.[44] In the expanded scaling, the anomalous Hall resistivity is expressed as

$$\rho_{AH} = a'\rho_{XX0} + a''\rho_{XXT} + b\rho_{XXT}^2 \qquad (4)$$

where $\rho_{XX0}$ is the residual resistivity, $\rho_{XXT}$ is the phonon induced resistivity, $a'$, $a''$, and $b$ are related to impurity induced skew scattering, phonon induced skew scattering, and Berry curvatures, respectively. Here the residual resistivity was taken to be the resistance at 2 K. The anomalous Hall resistivity vs longitudinal resistivity curves are shown in the insets of Fig. 11(c.d). The contribution of skew scattering ($a'$, $a''$) is comparable to the intrinsic contribution ($b\rho_{XX0}$) for both the x=0.3 and x=0.5 films providing further support that the expanded scaling is necessary for CoTi$_{1-x}$Fe$_x$Sb at other concentrations of Fe. Beyond the magnitude of the coefficient increasing for the higher content film, a proportionally lower contribution of the impurity induced skew scattering can be observed for the x=0.3 film.

### D) Evidence of Superparamagnetism

The dramatic change in the magnetic moment as well as the MR of the CoTi$_{0.7}$Fe$_{0.3}$Sb around 70 K is suggestive of a phase transition. To investigate this further, additional magnetization vs temperature (M-T) curves for the x=0.2, 0.3, and 0.5 samples were obtained and are shown in Fig 12(a). For this measurement, a constant magnetic field of 300 Oe was applied along the [110] sample direction and the magnetization was measured as a function of temperature during warm-up for samples that were zero-field-cooled (ZFC) and 20 kOe field-cooled (FC), respectively. The x=0.2 and x=0.3 M-T curves in the ZFC condition exhibit a blocking phenomenon with a peak in magnetization at around 70-100K. In contrast, the x=0.5 sample shows no peak in the ZFC curve, which is consistent with normal ferromagnetic behavior. The low temperature splitting between the ZFC and FC curves seen for the x=0.2 and x=0.3 samples could originate from a ferromagnet to superparamagnet transition, which can be observed in inhomogeneous magnetic systems in which ferromagnetic clusters are distributed in a nonmagnetic matrix[45]. The blocking phenomenon observed is attributed to the freezing of the magnetization of the ferromagnetic clusters at low temperature due to their magnetic anisotropy. In $Fe_2MnAl$, a ZFC and FC splitting in M-T curves was attributed to antiferromagnetic pinning of ferromagnetic parts[42]. However, no evidence of an antiferromagnetic phase was observed in any of the MvT measurements.

Above the blocking temperature, a superparamagnetic material should display paramagnetic-like behavior, but with an important identifying signature. Hysteresis curves that have been corrected for the temperature dependence of the spontaneous magnetization should approximately superimpose when plotted against H/T[45]. The $CoTi_{0.7}Fe_{0.3}Sb$ sample was tested for superparamagnetism by collecting hysteresis curves at temperatures between 150 K and 400 K. Figure 12(b) shows the measured magnetization curves plotted as $M_s(T)/j_s = f(j_s H/T)$, where $j_s = M_s(T)/M_s(0)$ for 150 K, 300 K, 350 K, and 400 K. The inset shows the fit to the saturation magnetization which gives a spontaneous magnetization $M_s(0)$=2.6x10$^{-5}$ emu and a Curie temperature $T_c$=480 K for this sample. The normalized M vs H/T curves nearly overlay for the 300 K, 350 K, and 400 K temperatures suggesting that the transition range from the blocked to the superparamagnetic state is from ~ 100 K to 300 K. During this transition the magnetic susceptibility gradually changes the slope for temperature independent (blocked state) through $1/k_B T$ (anistropic state with easy axes aligned with the external field) to $1/3k_B T$ (completely isotropic superparamagnetic state with magnetization described by the Langevin function). The magnetic moment of the 300 K data is fit by $M = M_s L(x)$, where L(x) is the Langevin function, with $x = \mu H/k_B T$. Here it is assumed that the system is comprised of noninteracting and monodisperse particles. The simple Langevin fit results in a particle magnetic moment of µ≈6700 µ$_B$ which corresponds to approximately 1700 Fe atoms contributing 3.9 µ$_B$/Fe atom.

While the M-T and M-H data are suggestive of superparamagnetism, there are no indications of non-h-H phases in the RHEED, XRD, or XPS spectra. Additionally, the magnetic moment nearly follows the expected 4 µ$_B$/Fe atom from Slater-Pauling for Fe occupying the Ti site in the h-H structure suggesting that Fe-Fe clusters, which would contribute closer to 2.2 µ$_B$/Fe atom, are not present. Thus, it can be inferred that the observed superparamagnetic behavior can be attributed to non-homogeneous distribution of the Fe atoms, leading to Fe rich h-H phase (e.g. $CoTi_{1-x-\delta}Fe_{x+\delta}Sb$), within an Fe poor semiconducting matrix ($CoTi_{1-x+\delta}Fe_{x-\delta}Sb$). This demixing was predicted for $CoTi_{1-x}Fe_xSb$ for the majority of intermediate compositions[46,47]. While no evidence of phase separation was observed in x=0.1 bulk crystals in transmission electron microscopy[16], the expected contrast would be quite weak, even for nanoparticles of CoFeSb within a CoTiSb matrix. Moreover, x=0.1 may still be outside the region of spinodal decomposition.

### E) Nanometer scale structural characterization

To investigate the nanometer scale distribution of Fe in the samples, atom probe tomography (APT) was performed on a x=0.3 film[48,49]. A 130 nm thick $CoTi_{0.7}Fe_{0.3}Sb$ film was grown for the purpose of the analysis with an *in-situ* deposited Ni capping layer of ~5nm used to prevent oxidation. An additional 150 nm of Ni was electron-beam deposited *ex-situ* on the samples to protect the regions of interest during the APT specimen preparation. Sharp tips were prepared with a FEI Helios 600 dual beam Focused Ion Beam (FIB) instrument following standard procedure with final FIB voltage down to 2 kV to minimize Ga induced damage[50]. APT analyses were performed with a Cameca 3000X HR Local Electrode Atom Probe (LEAP) operated in voltage-pulse mode with a sample temperature of 75 K to reduce the probability of tip fracture. A pulse fraction of 25% pulse to base voltage was chosen with a detection rate set to 0.005 atoms/pulse[49]. The APT 3D reconstruction was carried out using commercial software IVAS™. The reconstruction is optimized to visualize flat atomic planes in the Z-direction with the correct corresponding distance between planes[51].

Figure 13(a) are 25x25x50 nm³ 3D reconstructions of the $CoTi_{0.7}Fe_{0.3}Sb$ layer showing the four different elements. The measured elemental composition by APT are 35% of Co, 31% of Sb, 24% of Ti and 10% of Fe which is in good agreement with the expected 33% of Co, 33% of Sb, 24% of Ti and 10% of Fe. The small discrepancy between the measured and expected compositions may not be materials related but could be caused by the difficulty to adjust the APT evaporation parameters in a way that all elements are correctly detected[52]. In the 3D reconstructions, Co, Sb, and Ti rich clusters could not be visually directly identified. A homogenous distribution of these elements is observed. However, Fe rich and poor regions can be directly identified on the reconstruction. A statistical analysis of the data has been carried out in order to confirm the presence of possible Fe rich phases[53,54]. Figure 13(b) shows the radial distribution function (RDF) curves compared to Fe atoms. In these graphs, the ratio of the composition in shells drawn around each of the Fe atoms divided by the average composition in the sample is plotted versus the shell radius. The self-correlation curve is generated by measuring the composition of Fe (Fe-Fe) while the cross-correlation curve is generated by measuring the compositions of Co, Ti and Sb (Fe-Co, Fe-Ti and Fe-Sb). The RDF analysis of a homogeneous material would result in the self and cross-correlations curves being horizontal lines with a value of 1. This behavior is observed for the Fe-Co and Fe-Sb curves which indicates a homogenous distribution of Co and Sb around Fe atoms. However, a clear positive interaction (curve above 1) below 20 Å is found in the Fe-Fe curve showing that Fe-rich domains are present in the sample. This positive interaction corresponds to a negative interaction (curve below 1) in the Fe-Ti curve.

As expected from the crystal structure of the CoTi$_{0.7}$Fe$_{0.3}$Sb layer, a local Fe rich domain corresponds to a local depletion in Ti, further evidence Fe and Ti occupy the same crystallographic site in the h-H crystal structure. Self-correlation and cross-correlation curves were also plotted relatively to Co, Ti and Sb centers. Positive and negative interactions were not as clearly observed as in Fig. 13(b) suggesting more uniform distribution across the sample for Co, Ti, and Sb.

The Fe rich domains observed in APT of a CoTi$_{0.7}$Fe$_{0.3}$Sb film are consistent with the superparamagnetic behavior observed. The absence of pure nanoparticles within the film explains deviations away from traditional superparamagnetism. While APT analysis was not performed on other Fe content films, it can be expected that Fe rich regions likely exist in other composition films. The degree of Fe clustering will likely be a function of the Fe content of the film as well as the film growth/annealing temperatures. This may also partially explain the need for lower growth temperatures to achieve high quality, smooth films as determined from RHEED and XRD for higher Fe content films. Finally, the observed thermally activated behavior in the sheet resistance could be understood to originate from the non-homogenous Fe distribution that leading to a hopping-like transport instead of band transport. Future studies will examine the nanometer scale structural properties of other Fe content films and their dependence on growth/annealing temperatures.

## IV- Summary

In summary, epitaxial thin films of the substitutionally alloyed series CoTi$_{1-x}$Fe$_x$Sb and Co$_{1-y}$Fe$_y$TiSb were grown by MBE for concentrations $0.0 \leq x \leq 1.0$ and $0.0 \leq y \leq 0.5$. Fe concentration plays a significant role in determining the electrical and magnetic properties depending on which atomic site it substitutes. When Fe substitutes on the Ti site, the magnetic moment scales linearly with Fe content up to x=0.5 as ~3.9 $\mu_B$/Fe atom with a transition from weak to strong interaction as Fe content is increased. In contrast, a drastically reduced moment of ~0.4 $\mu_B$/Fe atom is observed when Fe substitutes for Co. Semiconducting-like transport can be observed for x≤0.5 with a strong anomalous Hall effect observed for the higher Fe content films that further supports the expanded scaling in the CoTi$_{1-x}$Fe$_x$Sb thin films. These tunable magnetic properties as well as simultaneous high resistance make CoTi$_{1-x}$Fe$_x$Sb thin films attractive for spintronic applications. Finally, the observed superparamagnetic behavior and APT analysis suggests Fe compositional fluctuations are present. These nano-scale compositional variations may be present in other quatenary alloyed Heusler compounds but with subtler effects. Interestingly, the Fe compositional fluctuations would likely be effective at scattering phonons, making these films a promising direction for CoTiSb based thermoelectrics.

**Acknowledgements**

The growth and transport measurements were supported by the Office of Naval Research through the Vannevar Bush Faculty Fellowship under award number N00014-15-1-2845. The theoretical work and the magnetic measurements at UCSB were supported by the U.S. Department of Energy under award number DE-SC0014388. We also acknowledge the use of facilities within the National Science Foundation Materials Research Science and Engineering Center (DMR 11–21053) at the University of California at Santa Barbara, the LeRoy Eyring Center for Solid State Science at Arizona State University, and Nanomagnetism and Spintronic Research Center (DPT under 2009K-120730) at Gebze Technical University. The calculations made use of the Extreme Science and Engineering Discovery Environment, NSF grant number ACI-1053575, and the high-performance computing and the Information Technologies resources at the University of Delaware. S.D.H. was supported in part by the NSF Graduate Research Fellowship under Grant No. 1144085.

**Figures**

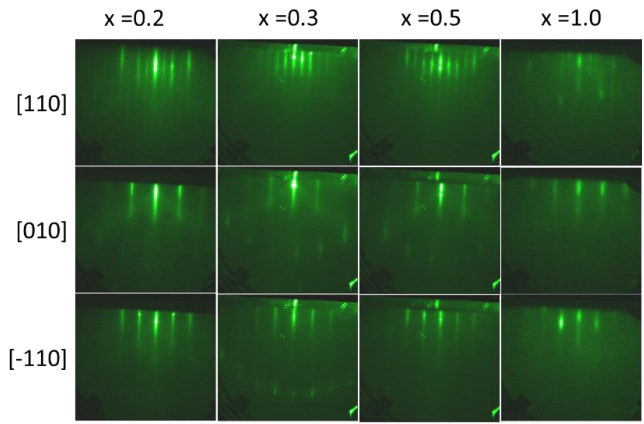

FIG 1. RHEED patterns of $CoTi_{1-x}Fe_xSb$ for x=0.2, 0.3, 0.5 and 1.0 along the [110], [010], and [1$\bar{1}$0] azimuths respectively. A clear (2x1) surface reconstruction in observed for x≤0.5 similar to that seen in pure CoTiSb.

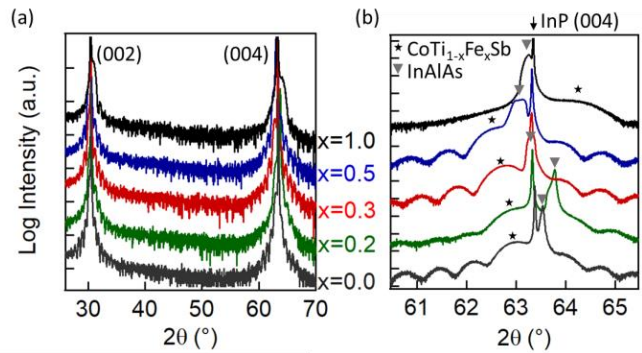

FIG 2. XRD ω-2θ scans for the $CoTi_{1-x}Fe_xSb$ films for x=0.2, 0.3, 0.5 and 1.0 grown on InAlAs/InP(001). (a) Survey scan along (00l) direction. (b) Close up of the (004) reflection.

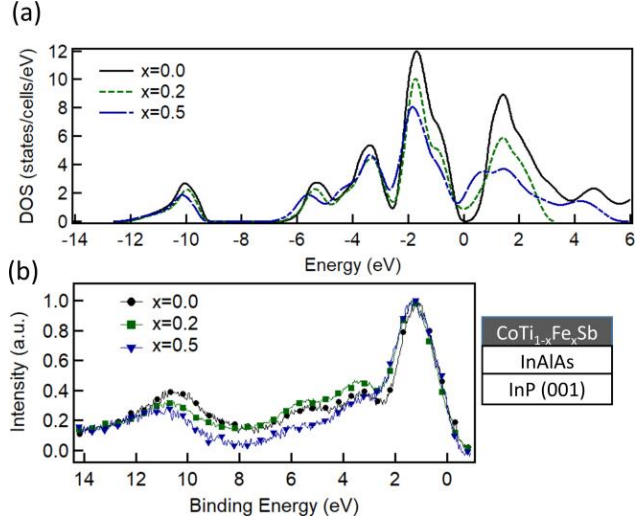

FIG 3. (a) Calculated total density of states plot for $CoTi_{1-x}Fe_xSb$ for x = 0.0, 0.2 and 0.5. (b) Normalized valence band XPS spectra collected for $CoTi_{1-x}Fe_xSb$ with x=0.0, 0.2, and 0.5 excited by Al Kα radiation.

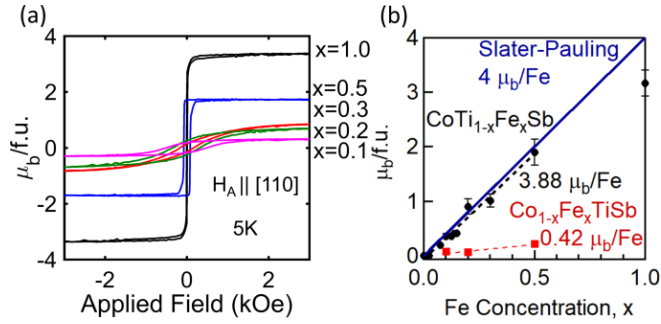

FIG 4. (a) SQUID magnetic hysteresis curves of 15 nm thick $CoTi_{1-x}Fe_xSb$ films for x=0.1, 0.2, 0.3, 0.5, and 1.0 at 5 K with the magnetic field applied along the [110] sample direction. (b) Magnetic moment per f.u. dependence on Fe concentration. Solid black circles and red squares are data points for $CoTi_{1-x}Fe_xSb$ and $Co_{1-y}Fe_yTiSb$ films respectively. Solid blue, dashed black, and dashed red lines correspond to the Slater-Pauling predicted 4 $\mu_B$/Fe atom, linear fit to the $CoTi_{1-x}Fe_xSb$ data, and linear fit to the $Co_{1-y}Fe_yTiSb$ data respectively.

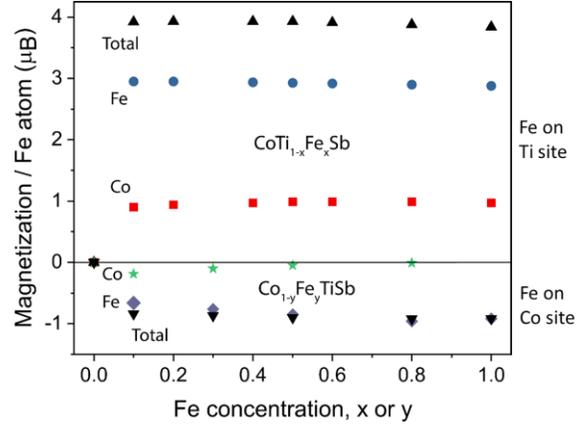

FIG 5. Calculated contributions to the total magnetization per Fe atom as a function of Fe concentration in $CoTi_{1-x}Fe_xSb$ and $Co_{1-y}Fe_yTiSb$ alloys. Note that the contribution from all the Co (or Fe) atoms are added up and divided by the number of Fe atoms in the supercell, with the largest contributions from the Co atoms sitting next to an Fe. The contributions from Ti and Sb atoms are negligible, i.e., less than 0.05 $\mu_B$ in magnitude, for all Fe concentrations.

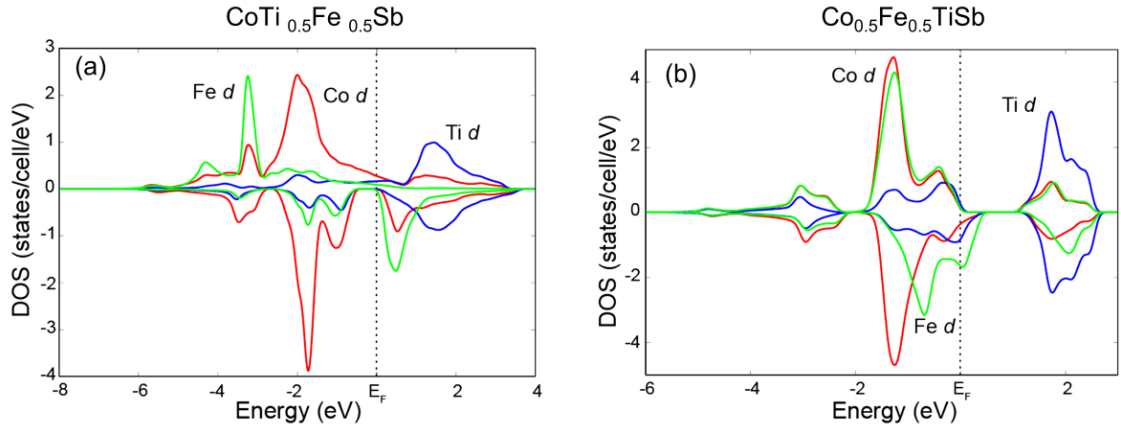

FIG 6. Projected Density of States for (a) $CoTi_{0.5}Fe_{0.5}Sb$ and (b) $Co_{0.5}Fe_{0.5}TiSb$ on Co, Fe and Ti d orbitals

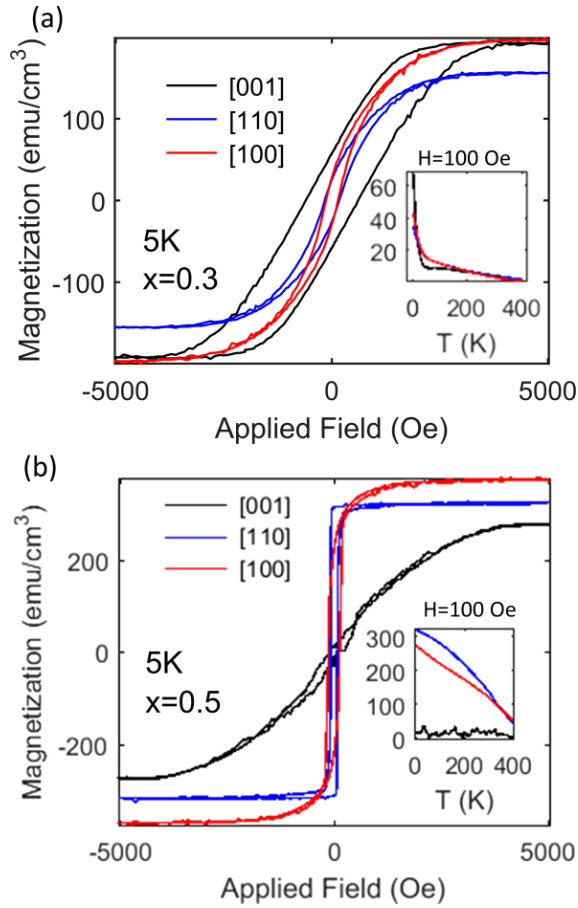

FIG 7. Magnetization hysteresis loops for (a) $CoTi_{0.7}Fe_{0.3}Sb$ and (b) $CoTi_{0.5}Fe_{0.5}Sb$ with the applied magnetic field along different crystallographic directions at 5K. [110] and [100] are in-plane directions while [001] is out-of-plane. The insets show the temperature dependence of the magnetic moment between 5 and 400 K with 100 Oe applied field.

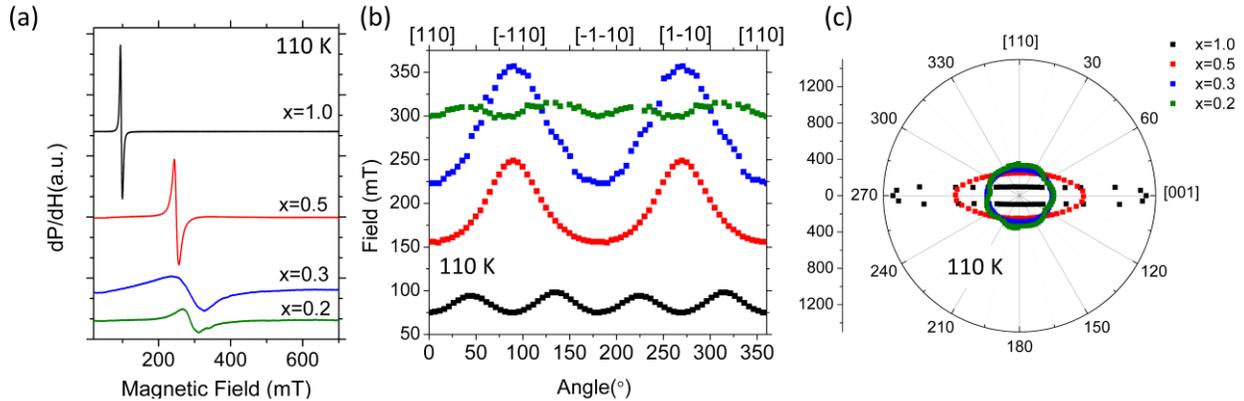

FIG 8. (a) In-plane FMR spectra for the $CoTi_{1-x}Fe_xSb$ alloy series for x=0.2, 0.3, 0.5, and 1.0. Angular dependence of FMR field observed at 110K for $CoTi_{1-x}Fe_xSb$ films for (b) in-plane measurement geometry in Cartesian coordinates and (c) out-of-plane in polar coordinates.

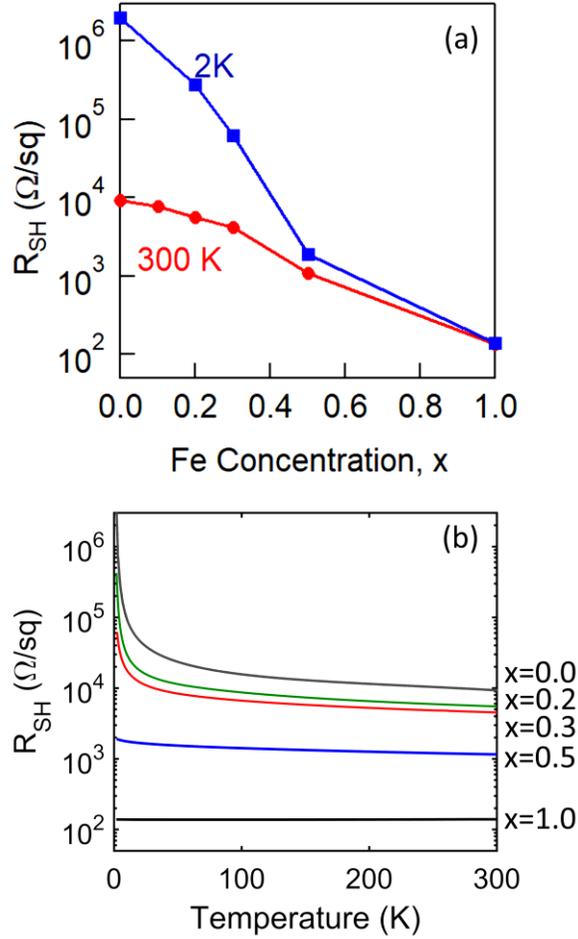

FIG 9. Longitudinal sheet resistance ($R_{SH}$) measurements for ~15 nm thick $CoTi_{1-x}Fe_xSb$ films x=0.0, 0.2, 0.3, 0.5, and 1. (a) The 2 K and 300 K $R_{SH}$ as a function of Fe concentration and (b) $R_{SH}$ as function of temperature.

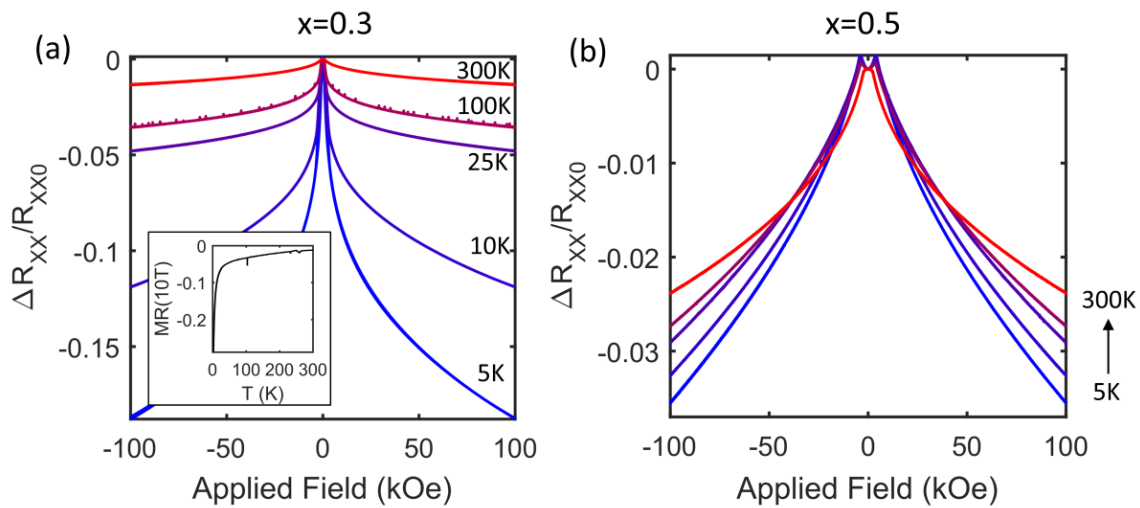

FIG 10. Temperature dependence of the longitudinal MR curves for (a) $CoTi_{0.7}Fe_{0.3}Sb$ and (b) $CoTi_{0.5}Fe_{0.5}Sb$ with the applied magnetic field out of plane for 5, 10, 25, 100, and 300 K. The MR is defined as $(R_{XXH}-R_{XX0})/R_{XX0}$ where $R_{XXH}$ and $R_{XX0}$ are the longitudinal resistances measured at external magnetic field and zero magnetic field respectively. The inset of (a) shows the MR as a function of temperature at 100 kOe.

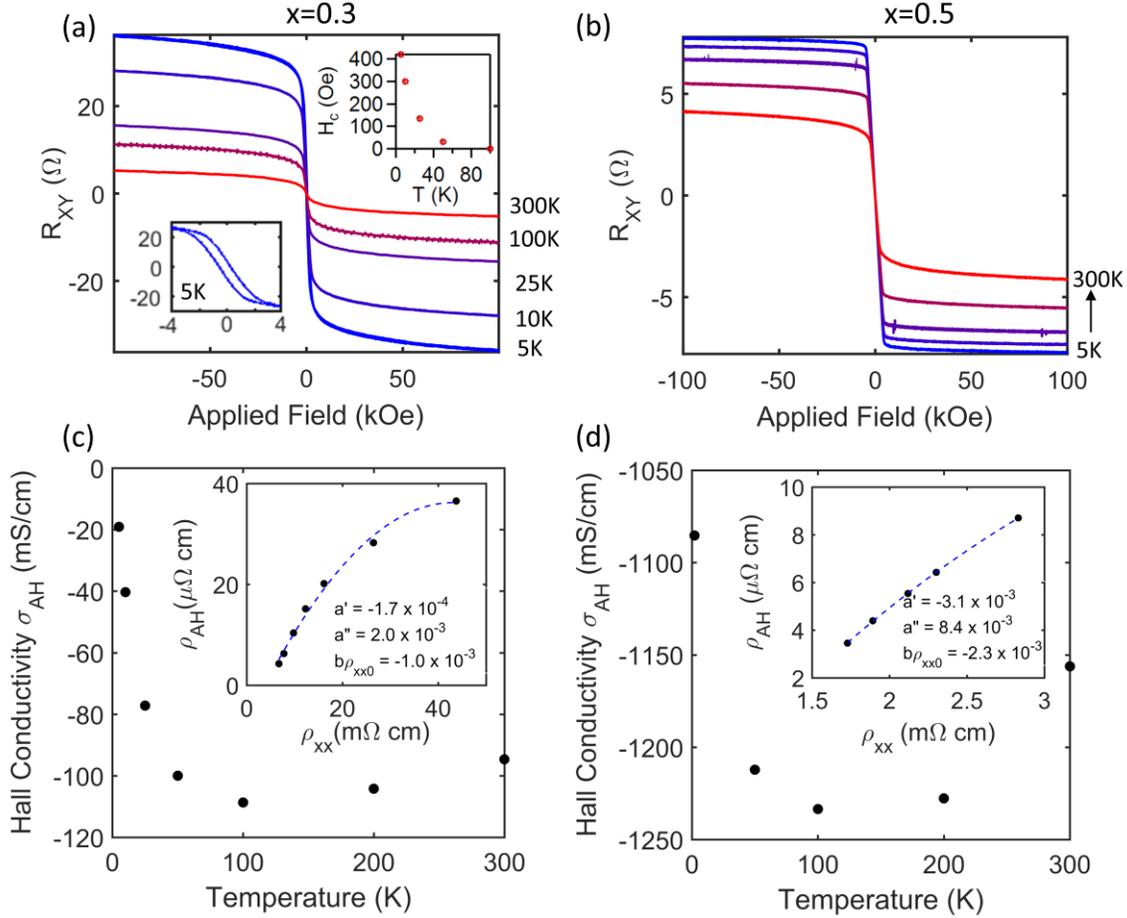

FIG 11. Anomalous Hall effect curves for 15 nm thick (a) $CoTi_{0.7}Fe_{0.3}Sb$ and (b) $CoTi_{0.5}Fe_{0.5}Sb$ with the applied magnetic field out of plane for 5, 10, 25, 100, and 300 K. The insets of (a) show the observed coercive field at 5K and the temperature dependence. Hall conductivity vs temperature curves for (c) $CoTi_{0.7}Fe_{0.3}Sb$ and (d) $CoTi_{0.5}Fe_{0.5}Sb$. The insets of (c) and (d) show the anomalous Hall resistivity vs longitudinal resistance curves fitted by the expanded scaling expression.

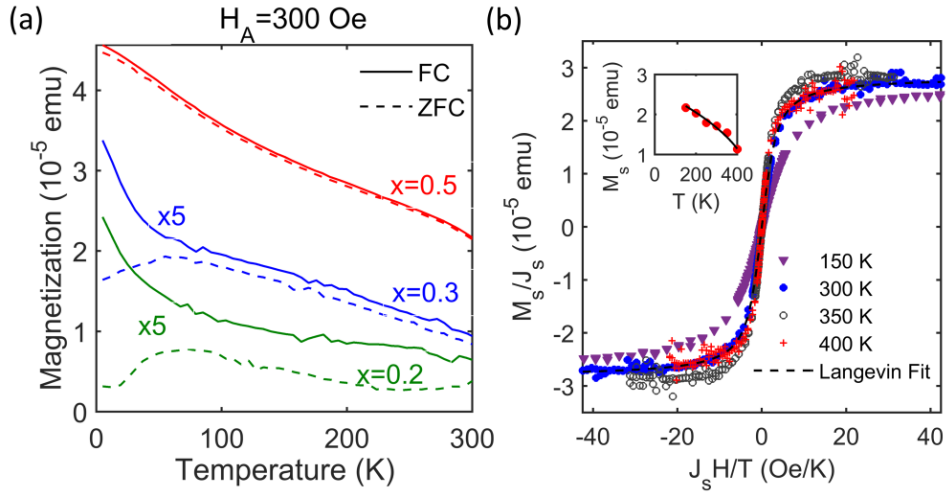

FIG 12. (a) Temperature dependence of the magnetization for the x=0.2, 0.3, and 0.5 samples applied field $H_A$=300 Oe. Curves were taken while warming up in the ZFC and FC conditions with 0 and 20 kOe applied field respectively during cool down. (b) Normalized magnetization as a function of $j_s H/T$ at temperatures of 150, 300, 350, and 400 K for the x=0.3 sample. The dashed line is a fit of the 300 K data to the Langevin function. The inset shows the temperature dependence of the saturation magnetization and resulting fit used to determine $j_s(T)$.

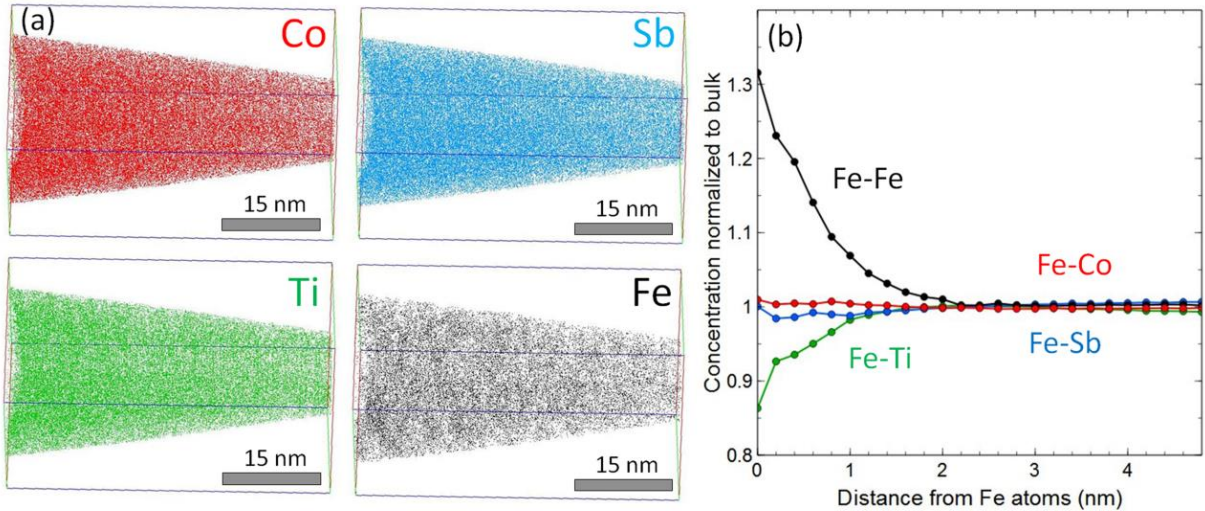

FIG 13. (a) Three-dimensional (3D) reconstruction of the atomic distribution from atom probe tomography performed on a x=0.3 film for Co, Ti, Fe, and Sb. The colors red, green, blue, and black correspond to Co, Ti, Sb, and Fe respectively. (b) Radial distribution function (RDF) curves for Fe-Fe, Fe-Co, Fe-Sb, and Fe-Ti.

## Tables

| Stoichiometry | No. of Valence Electrons | Slater Pauling (in $\mu_B$ / Fe) | Calculated Magnetic moment (in $\mu_B$ / Fe) |
|---|---|---|---|
| $CoTi_{1-x}Fe_xSb$ | $18+4x$ | 4.0 | 3.92 |
| $Co_{1-y}Fe_yTiSb$ | $18-y$ | -1.0 | -0.90 |
| $Co_{1-y}Ti_{1-x}Fe_{x+y}Sb$ (x=y) | $18+3(x+y)$ | 1.5 | 1.48 |
| $(Co_{1-y}Fe_y)(Ti_{1-x}Co_x)Sb$ (x=y) | $18+4y$ | 4.0 | 2.38 |

Table I. Comparison between calculated magnetic moment from density functional theory and expected magnetic moment from Slater Pauling rule (in µB / Fe) for different stoichiometries/disorders of Fe substitution in CoTiSb